\documentclass{elsart4}

\usepackage{amssymb}

\usepackage{graphics}
\usepackage{graphicx}
\usepackage{epsfig}
\usepackage{amssymb}
\journal{Physica B}

\newcommand{\be}{\begin{equation}}
\newcommand{\ee}{\end{equation}}
\newcommand{\ba}{\begin{eqnarray}}
\newcommand{\ea}{\end{eqnarray}}

\begin{document}

\begin{frontmatter}


 \title{Relation Between Local Temperature Gradients and the Direction of Heat Flow in Quantum Driven Systems}
 \author{Alvaro Caso, Liliana Arrachea and Gustavo S. Lozano\corauthref{cor1}}
 \address{Departamento de F\'isica, FCEyN, Universidad de Buenos
   Aires, Pabell\'on 1, Ciudad Universitaria, 1428 Buenos Aires, Argentina}
 \corauth[cor1]{email: lozano@df.uba.ar}





\begin{abstract}
We introduce thermometers to define the local temperature of an
electronic system driven out-of-equilibrium by local ac fields. 
We discuss the behavior of the local temperature along the
sample, showing that it exhibits spatial fluctuations following an
oscillatory pattern. 
We show explicitly that the local temperature is the correct indicator
for heat flow.  
\end{abstract}

\end{frontmatter}

\section{Introduction}
Heat transport at the meso and nano scale is subject of wide
increasing interest at present and hence it is subject of intense
research. 
There are basically two motivations for this. 
On the one hand, the technological trend towards miniaturization of
electronic circuits pushes for a better understanding of the
mechanisms for heat production and energy flow at the microscopic
level. 
On the other hand, from a more general point of view, one is often
faced with situations in which the very fundamental concepts of
standard Statistical Mechanics and Thermodynamics are put into test.  
This is for instance the case when the system under consideration is
driven out of equilibrium. 

In a recent work, \cite{cal} we have addressed the issue of
identifying effective temperatures in the context of transport in
electronic quantum systems driven out of equilibrium by external
(periodic) pumping potentials. 
Examples of this type of systems are quantum dots with ac voltages 
acting at the walls (quantum pumps). 
These systems can not only dissipate energy in the form of heat, but
can also pump energy between the different reservoirs, generating
refrigeration.  
We have defined a ``local'' temperature along these set ups by
introducing a thermometer, i.e. a macroscopic system which is in local
equilibrium with the system, even when the system itself is out of
equilibrium. 
This is the thermal analogue of voltage probe discussed in Ref.
\cite{fourpoint,fourpointfed}. 

The aim of this work is to further investigate the concept of local
temperature in quantum driven systems. In particular, our goal is to
show that indeed such parameter signals the direction of heat flow,
acting as a {\em bona fide} temperature. 

\section{Model and theoretical treatment}
Our setup, including the device with the driven system in contact to
reservoirs and thermometer is described by the  Hamiltonian: 
\ba
H(t)&=& H_{sys}(t)+H_{cP}+H_{P},\nonumber\\
H_{sys}(t)&=& H_{L}+H_{cL}+ H_{C}(t)+H_{cR}+ H_{R}. 
\ea 
The piece $H_{sys}(t)$ contains the term describing the central system
($C$) with the AC fields, $H_C(t)=H_0+H_V(t)$ as well as terms
corresponding to  left ($L$)  and right ($R$) reservoirs with the
ensuing contacts $H_{cL}$ and $H_{cR}$. 
The term $H_P$ represents the thermometer. 
It consists in a macroscopic system weakly coupled to a given point
$lP$ of $C$, through a contact described by $H_{cP}$. 
It behaves like a reservoir with a temperature $T_{lP}$ that is
determined by the condition of a vanishing heat flow between it and
$C$. 
This is the thermal counterpart of a voltage probe (see Ref.
\cite{fourpoint,fourpointfed}). 
All the reservoirs are modeled by systems of non-interacting electrons
with many degrees of freedom:  
$H_{\alpha}= \sum_{k \alpha } \varepsilon_{k \alpha} c^{\dagger}_{k
  \alpha} c_{k \alpha}$, being $\alpha=L,R,P$. The corresponding
contacts are $H_{c \alpha}= w_{c \alpha} (c^{\dagger}_{k \alpha} c_{l
  \alpha} +c^{\dagger}_{l \alpha} c_{k \alpha})$, where $l \alpha$
denotes the coordinate of $C$ at which the reservoir $\alpha$ is
connected. 
We take into account the non-invasive property of the thermometer
\cite{fourpoint} by treating $w_{cP}$ at the lowest order of
perturbation theory when necessary. 
We leave for the moment $H_C$ undetermined as much of the coming
discussion is model independent. 

The dynamics of the system is best described within the
Schwinger-Keldysh Green functions formalism. 
This involves the calculation of the Keldysh and retarded Green
functions, 
\begin{eqnarray}
G^{K}_{l,l'}(t,t')&=& i \langle c^{\dagger}_{l'}(t') c_l(t) - c_l(t)
c^{\dagger}_{l'}(t') \rangle, \nonumber \\
G^R_{l,l'}(t,t')&=& -i \Theta(t-t') \langle c_l(t)
c^{\dagger}_{l'}(t')+c^{\dagger}_{l'}(t') c_l(t) \rangle ,
\label{green}
\end{eqnarray}
where the indexes $l,l'$ denote spatial coordinates of the central
system. 
These Green functions can be evaluated after solving the Dyson
equations. 
For ac driven systems, it is convenient to use the Floquet Fourier
representation of these functions \cite{liliflo}: 
\begin{equation}
 G_{l,l'}^{K,R}(t, t-\tau) = \sum_{k=-\infty}^{\infty}
 \int_{-\infty}^{\infty} \frac{d\omega}{2 \pi}
 e^{-i (k \Omega_0 t + \omega \tau)}  G_{l,l'}^{K,R}(k,\omega),
\end{equation}
where $\Omega_0$ is the frequency of the ac fields.

\section{Defining the local temperature}
As we did in Ref. \cite{cal} we define the local temperature as the
value of $T_P$ (i.e the temperature of the probe) such that heat
exchange between the central system and the probe vanishes.   
As shown in \cite{liliheat}, given $H_C(t)$ without many-body
interactions, the heat current from the central system to the
thermometer can be expressed as ($\hbar=k_B=1$)
\ba
& &J_P^Q (T_{lP})= \sum_{\alpha=L,R,P} \sum_{k=-\infty}^{\infty}
\int_{-\infty}^{\infty} \frac{d\omega}{2 \pi} 
 \{  [f_\alpha(\omega)-f_P(\omega_k)] \nonumber \\
& &   \times (\omega_k - \mu)
\Gamma_P(\omega_k) \Gamma_\alpha(\omega) \left|
G^R_{lP,l\alpha}(k,\omega)\right|^2 \} \label{jq} 
\ea
where $\omega_k=\omega+k\Omega_0$, $\Gamma_{\alpha}(\omega) = -2 \pi
|w_{\alpha}|^2 \sum_{k \alpha} \delta(\omega-\varepsilon_{k \alpha})$ 
is the spectral function that determines the escape to the reservoir
$\alpha$, and $f_\alpha(\omega)= 1/[e^{\beta_{\alpha}(\omega
    -\mu)}+1]$, is the Fermi function, which depends on the
temperature $T_{\alpha}=1/\beta_{\alpha}$ and the chemical potential
of the reservoir $\alpha$.
Thus, the local temperature $T_{lP}$ corresponds to the solution of
the equation $J_P^Q(T_{lP}) = 0$. In general, the exact solution must
be found numerically, however, an exact analytical expression can be
obtained within the weak-coupling and low-temperature $T$
regime.\cite{cal} 

\section{Results}

In this section we present results for a central device consisting of
non-interacting electrons in a one-dimensional lattice:
\be
H_0= -w \sum_{l,l^{\prime}} (c^\dagger_{l} c_{l^{\prime}} + H. C.),
\ee
where $w$ denotes a hopping matrix element between neighboring
positions $l,l^{\prime}$ on the lattice, and a driving term of the
form: 
\be \label{hv}
 H_V(t)= (\varepsilon_{l1} + e V(t) ) c^\dagger_{l1} c_{l1} ,
\ee
with $V(t)= V_0 \cos( \Omega_0 t)$, being $l1$ the position at where
the local ac field is applied.

\begin{figure}
\includegraphics[width=7.5cm]{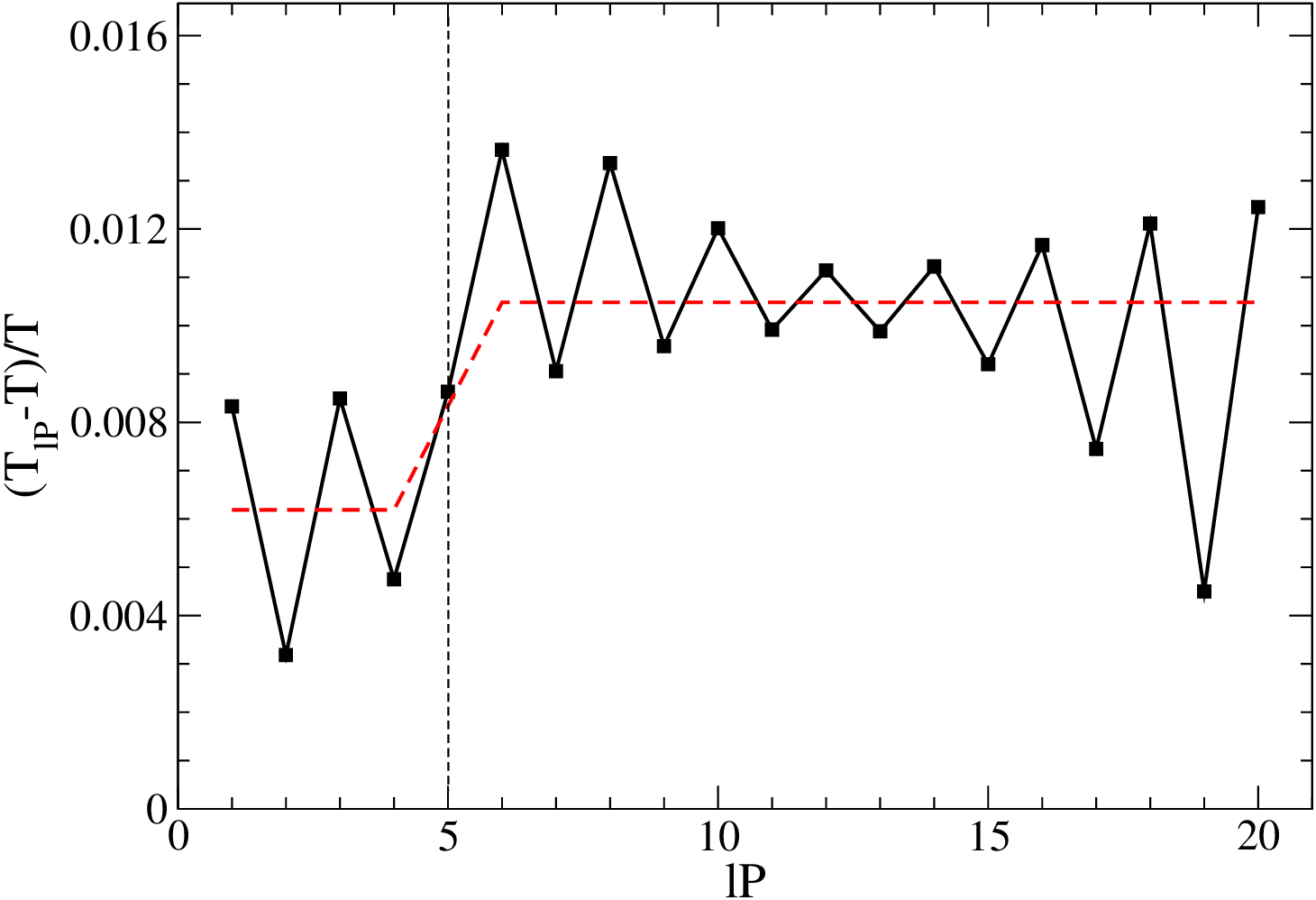}
\caption{(color online) Local (black solid) and coarse grained (red
  dashed) temperatures along a one-dimensional  model of N=20 sites
  with one ac field at the position indicated in dotted line. The
  system is in contact with two reservoirs with chemical potentials
  $\mu=0.2$ and temperature $T=0.03$. The driving frequency is
  $\Omega_0=0.1$, the amplitude is $V_0=0.05$ and
  $\varepsilon_{l1}=1$.}  
\label{fig1}
\end{figure}

In Fig. \ref{fig1} we show that the local temperature varies along the
sample. As we discused in Ref. \cite{cal}, the local temperature
presents oscillations modulated by 2$k_F$, being $k_F$ the Fermi
vector of the electrons leaving the reservoirs. This oscillations are
due to the coherent transport of the electrons through the device and
account for scattering processes at the point of the structure where
the ac voltage is applied.

\subsection{Coarse-grained temperature and the direction for the heat
  flow}
We now turn  to explore the relation between this local temperature
and the heat flow.  
In order to do this, it is necessary to connect this local microscopic 
temperatures to coarse grained temperatures for the two thermal
regions in which is divided the sample. A left (l) region defined
between the $L$ reservoir and the point at which the ac voltage is
applied and a right (r) region defined between the point at which the
ac voltage is applied and $R$ reservoir.
Then, it is straightforward to define the mean temperature of these two
regions as follows:

\begin{equation}
T^{CG}_{\gamma} = \frac{1}{N_{\gamma}}  \sum_{l \subset \gamma}  T_{l},
\end{equation}
where $\gamma=$ l,r, while $N_{\gamma}$ denotes the number of points
contained in each of these regions. 
The mean temperatures for both regions are shown in Fig. \ref{fig1}.

We now turn to show that the difference between the coarse grained
temperatures completely determine the direction of the heat flow as
follows: 
\begin{equation}\label{kappa}
J^Q_{\alpha \rightarrow \gamma} = \kappa_\gamma (T^{CG}_{\alpha} -T^{CG}_{\gamma}),
\end{equation} 
where for the case of the reservoirs $T^{CG}_{\alpha} \equiv
T_{\alpha}$ , while $\kappa_\gamma$ is a thermal conductance which
might depend on temperature and some of the parameters of the system. 

\begin{figure}
\centering
\begin{tabular}{cc}
\includegraphics[width=3.7cm]{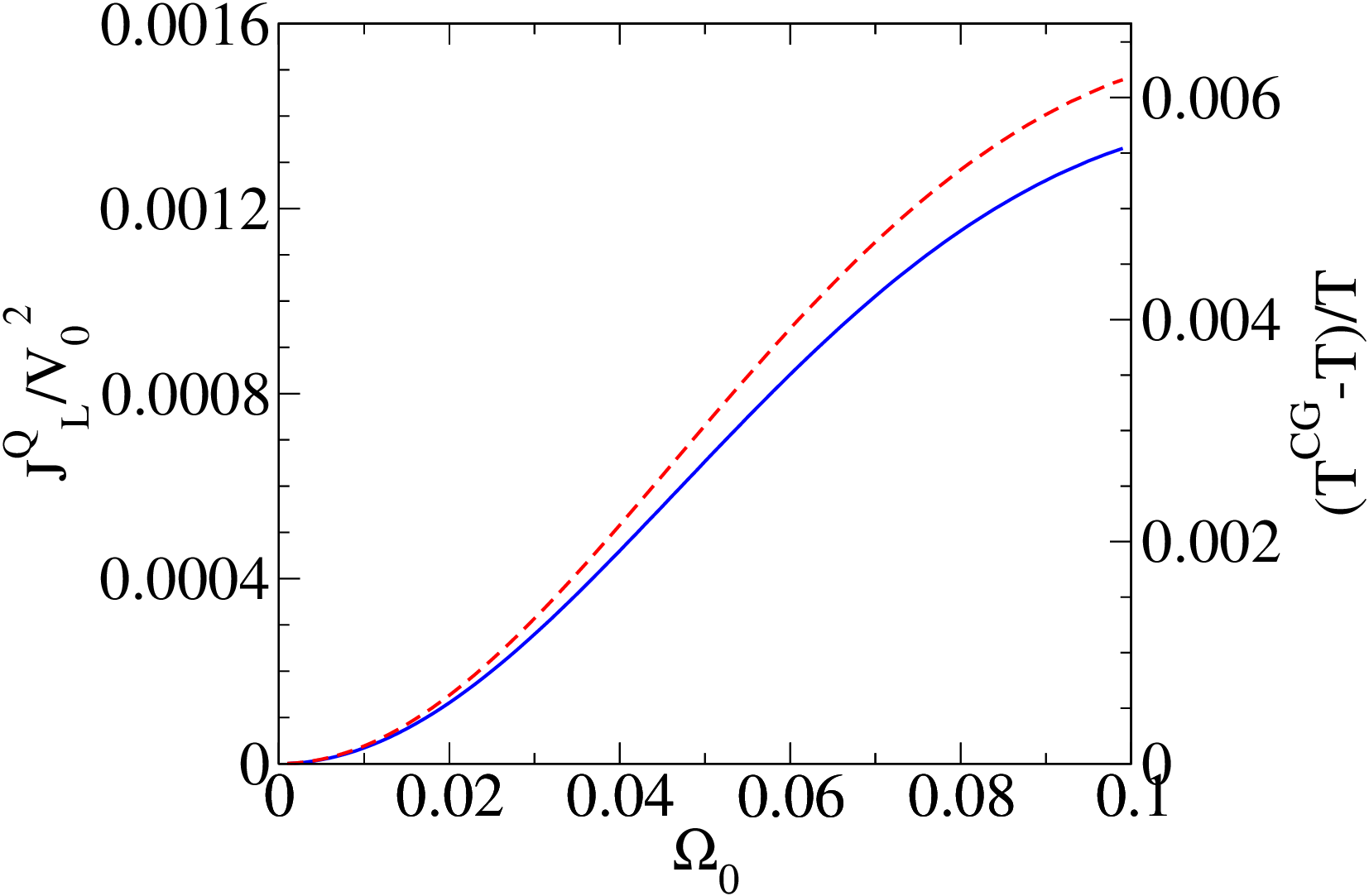} & 
\includegraphics[width=3.7cm]{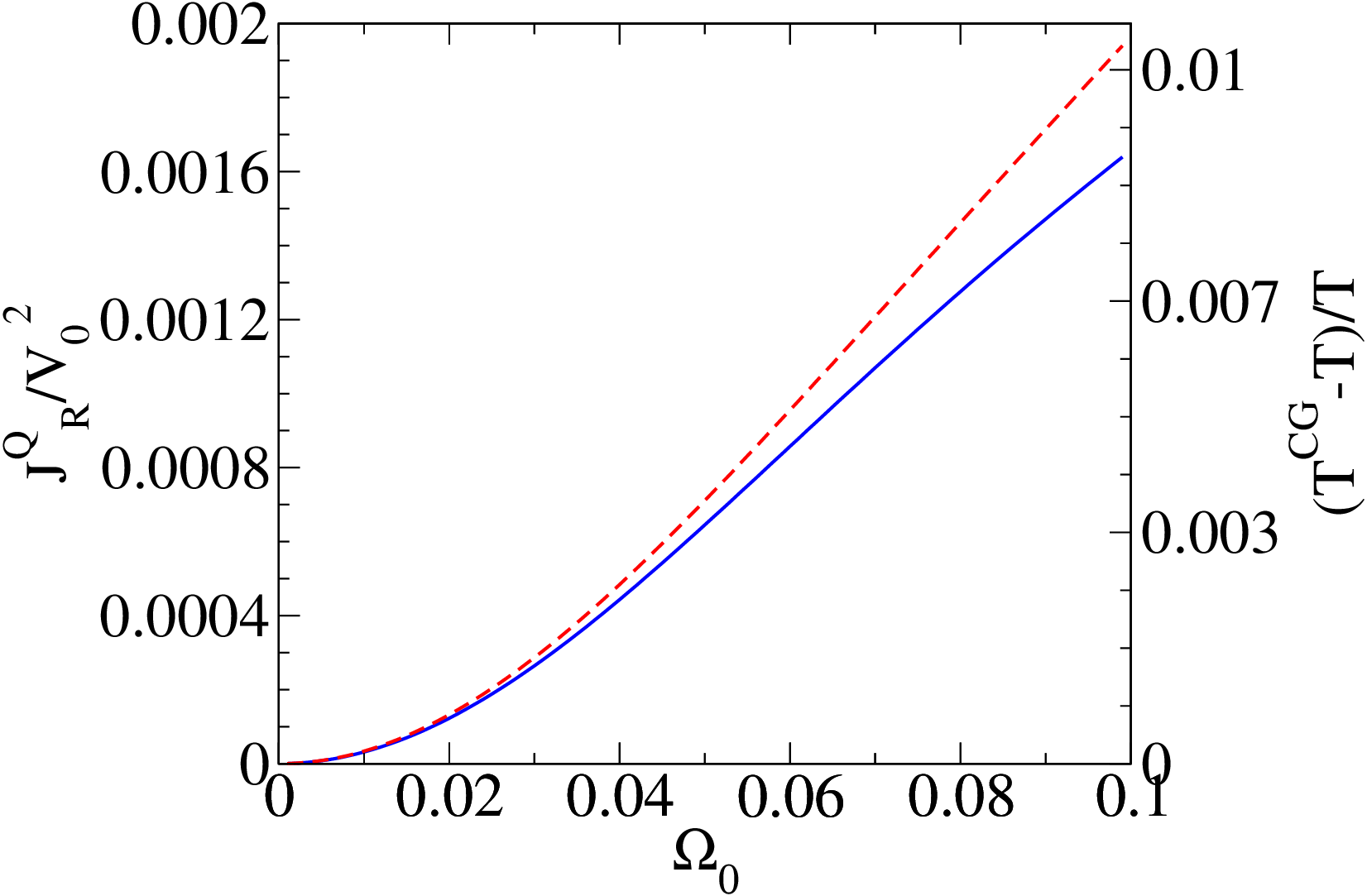} \\\\
\includegraphics[width=3.7cm]{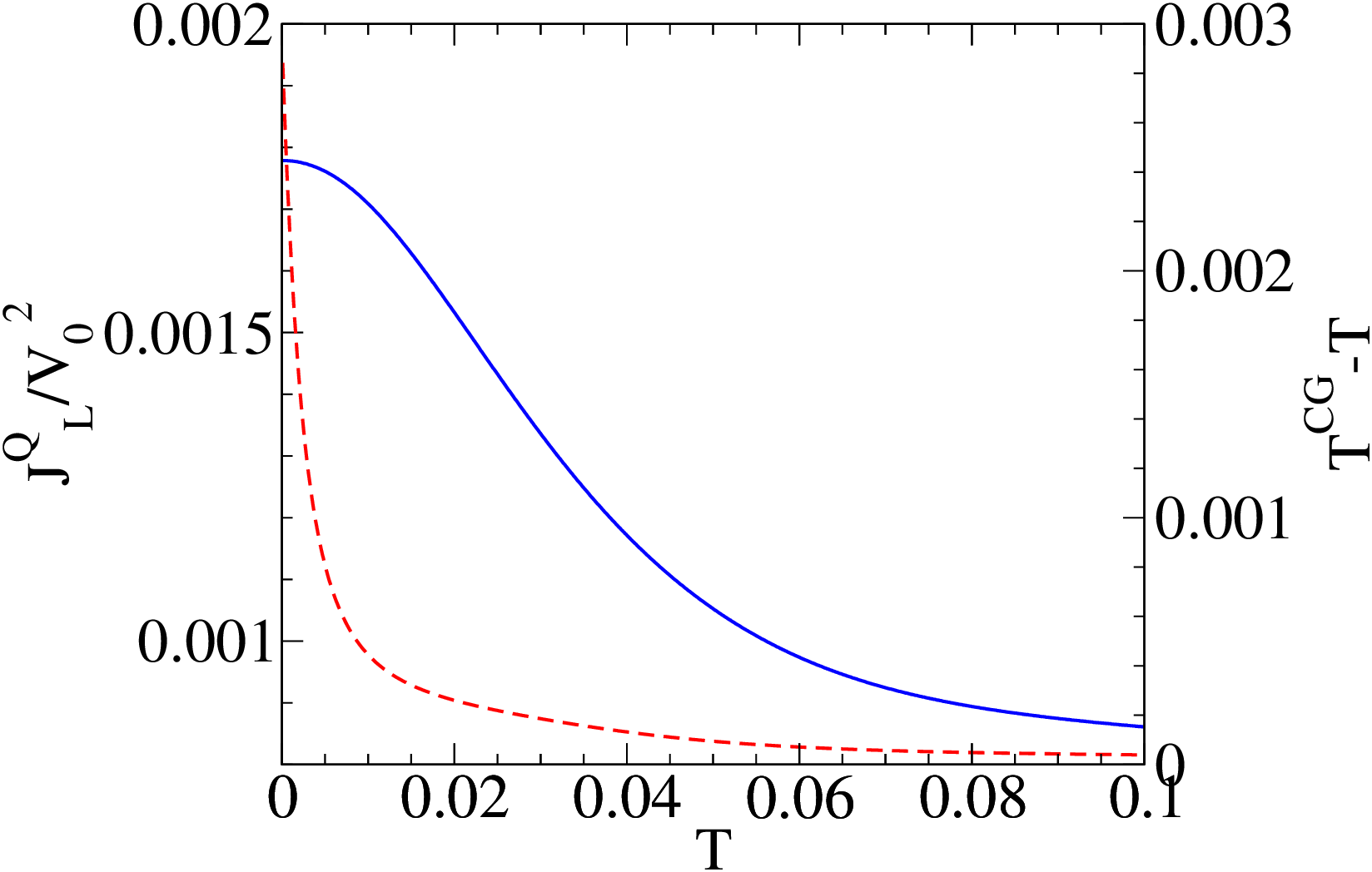} & 
\includegraphics[width=3.7cm]{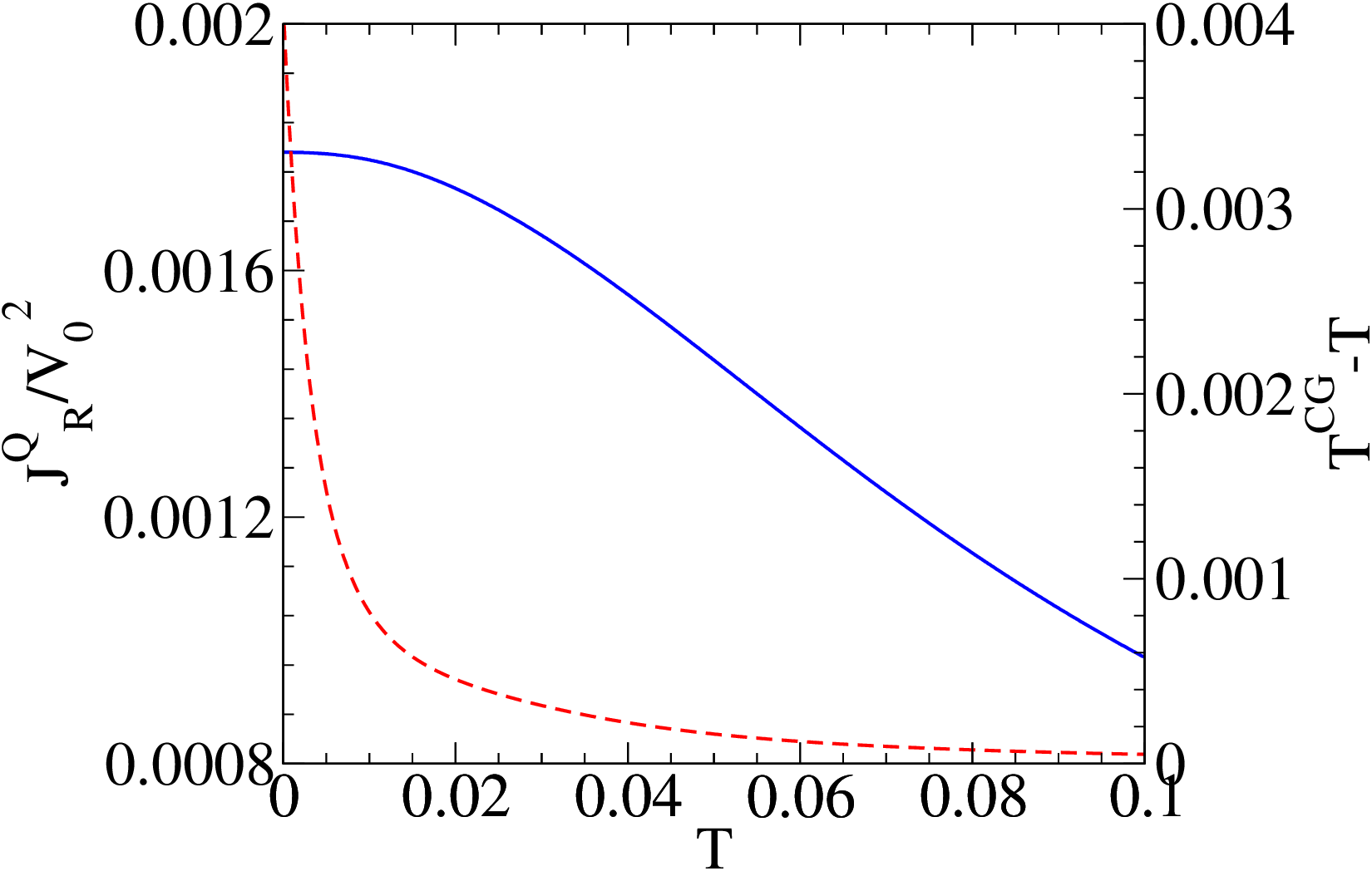}
\end{tabular}
\caption{(color online) Upper panel: heat flow (solid) and local
  temperature difference (dashed) between the system and the left
  reservoir (left figure) or the right reservoir (right figure) as a
  function of the driving frequency $\Omega_0$. The temperature of the
  reservoirs is $T=0.03$, the amplitude of the driving is $V_0=0.05$
  and $\varepsilon_{l1}=1$. The chemical potential of the reservoirs
  is $\mu=0.2$.
  Lower panel: heat flow (solid) and local temperature difference
  (dashed) between the system and the left reservoir (left figure) or
  the right reservoir (right figure) as a function of temperature $T$
  of the reservoirs. The driving frequency is $\Omega_0$=0.01, the
  amplitude is $V_0=0.05$ and $\varepsilon_{l1}=1$. The chemical
  potential of the reservoirs is $\mu=0.2$.  }
\label{fig2}
\end{figure}

Results are shown in Fig. \ref{fig2}, as a function of the
driving frequency $\Omega_0$ (upper panel) and the temperature $T$ of
the reservoirs (lower panel). 
The flow is defined positive when the heat flows to the reservoir. 
In the same figure, we show $T^{CG}_\gamma - T_\alpha$. 
This is indeed the expected situation in which the effective
temperature is larger than the leads temperature and we see explicitly
that the heat flow occurs as expected, from the ``hot'' sector (the
system) to the colder one (the leads). 
In a pumping regime we expect to find situations in which heat can be
extracted from one lead to pump it into the system, but this kind of
regime is not possible with only one ac field acting on the system. 

As we can see in the upper panel of Fig. \ref{fig2}, for increasing
driving frequency $\Omega_0$, the difference between the
coarse-grained temperature $T^{CG}_\gamma$ and the temperature $T$ of
the reservoirs increases, and so does the heat current that flows into
the reservoir. 
The proportionality constant between this two quantities is the
thermal conductivity $\kappa_\gamma$ ($\gamma=l,r$) that seems not to
depend on the driving frequency $\Omega_0$, but further study is
necessary in order to arrive to that conclusion. 

In the lower panel of Fig. \ref{fig2} we can see that for increasing
temperature $T$ of the reservoirs, the difference  $T^{CG}_\gamma - T$
decreases, and so does the heat current that flows into the
reservoir. 
In this case, the proportionality constant between this two quantities
does depend on temperature $T$. The temperature difference decreases
faster than the heat current so the thermal conductivity
$\kappa_\gamma$ is an increasing function of $T$, but further studies
are necessary in order to determine the dependence on $T$. 
The fact that $T^{CG}_\gamma - T$ goes to zero reflects the fact that
at a high temperature, the effect of the driving becomes washed up and
the sample becomes mainly heated due to the contact with a high
temperature environment. This is in complete agreement with previous
results obtained in a similar system. \cite{cal}

\section{Conclusion}
We studied the behavior of the local temperature in a quantum system
driven out of equilibrium by one ac field. This quantity indicates a
global heating of the sample, which manifests itself in the form of
mean temperatures higher than the one of the reservoirs. The occurrence
of $2k_F$ oscillations in the local temperature is an indication of
quantum interference, due to coherence in the energy propagation, in
complete agreement with the results obtained in a similar system (see
Ref. \cite{cal}).  
We showed that the difference between the mean temperature of the l
(r) region of the sample and the one of the reservoirs defines the
heat flow that enters the L (R) reservoir. 
In a pumping regime, we expect that the mean temperature of the region
in contact with the reservoir from which heat is extracted shall be
lower than the one of the reservoirs. 
This kind of regime is not possible in the studied system and will be
subject of future investigation. 

\section{Acknowledgments}
We acknowledge support from CONICET, ANCyT, UBACYT, Argentina and
J. S. Guggenheim Memorial Foundation (LA).



\end{document}